%% file: d0_zc_v2.tex
\RequirePackage{lineno}
\documentclass[aps,prd,twocolumn,superscriptaddress,groupedaddress,floatfix]{revtex4}

\usepackage{graphicx}
\usepackage{dcolumn}
\usepackage{bm}
\usepackage{amsmath,amssymb}   
\usepackage{subfigure}
\usepackage{multirow}
\usepackage{afterpage}
\usepackage{url}
\usepackage{latexsym}
\usepackage{lscape}
\usepackage{subfigure}
\usepackage{setspace}

\begin{document}



\hspace{4.8in} \mbox{FERMILAB-PUB-18-303-E }


\title{Evidence for  {\boldmath  $Z_c^{\pm}(3900)$} in semi-inclusive decays
of {\boldmath  $b$}-flavored  hadrons}

\input author_list.tex

\date{June 30, 2018}

\begin{abstract}
We present evidence for the exotic charged charmonium-like state
 $Z_c^{\pm}(3900)$ decaying to $J/\psi \pi^{\pm}$
in semi-inclusive weak decays of $b$-flavored hadrons. 
The signal is correlated with a  parent $J/\psi \pi^+ \pi^-$ system
in the invariant mass range 4.2$-$4.7~GeV that would  include   the exotic
structure $Y(4260)$.
The study is based on  $10.4~\rm{fb^{-1}}$ of $p \overline p $ collision
data  collected by the  D0 experiment at  the Fermilab Tevatron collider.
\end{abstract}

\maketitle


\section{Introduction}

The charged charmonium-like state $Z_c^{\pm}(3900)$ was discovered in 2013 simultaneously
by the  Belle~\cite{belle2013} and BESIII~\cite{bes2013}  collaborations in the sequential process 
$e^+e^- \rightarrow  Y(4260)$,  $Y(4260) \rightarrow Z_c^+(3900) \pi^-$, $Z_c^+(3900) \rightarrow J/\psi \pi^+ $
 (charge conjugate processes are
 implied throughout).
Their fits of the $Z_c^+(3900)$ signal with an $S$-wave Breit-Wigner signal shape
and an incoherent background gave the signal parameters
     $m=3894.5\pm6.6\pm4.5$~MeV, $\Gamma=63\pm34\pm26$~MeV
and  $m=3899.0\pm3.6\pm4.9$~MeV, $\Gamma=46\pm10\pm20$~MeV, respectively.
The $Z_c^+(3900)$ cannot be a conventional quark-antiquark meson
as it is charged and  decays via the strong interaction to charmonium. Its minimal quark content
is thus  $c\overline c u \overline d$.

Since the original observation,  the understanding of both the  $Z_c^+(3900)$ and $Y(4260)$  has evolved. 
The BESIII collaboration has measured~\cite{bes3y}  the $e^+e^- \rightarrow J/\psi \pi^+ \pi^-$ 
cross section at a range of energies from 3.77~GeV to 4.60~GeV
and reported that the  $Y(4260)$ may consist of two states: a narrow 
state at about 4.22~GeV and a wider one at about 4.32~GeV above a continuum
that may also be consistent with a broad resonance near 4.0~GeV.   
Currently  the  ``$Y(4260)$'' is believed to be  composed
of two states: a lower-mass  narrower state denoted by the Particle Data Group (PDG)~\cite{pdg}  as 
$\psi(4260)$ 
with  mass  $m=4230\pm8$~MeV and width  $\Gamma=55\pm19$~MeV and a higher-mass broader state 
$\psi(4360)$  with  $m=4368\pm13$~MeV and $\Gamma=96\pm7$~MeV.

The  $Z_c^+(3900)$ is  close in mass to $X(3872)$ and also close to the open-charm 
 $D^* \overline D$ threshold, so it  may be a ``molecular'' state composed of a loosely bound
pair of colorless,   quark-antiquark pairs containing a charm and a light quark $(c\bar d )$ and $(\bar c u)$,
 the isovector  analog of the  $X(3872)$.
A mass enhancement is also seen in the $D^* \overline D$ system~\cite{besdd}
but the fit for this channel gives a different mass and width compared to that for the $J/\psi \pi^+$
 channel.

The PDG~\cite{pdg}  assumes that it is a single   resonance decaying to two final states.
It lists   it as $Z_c(3900)$ with    $m=3886.6\pm2.4$ MeV
and $\Gamma=28.2\pm 2.6$ MeV.
The spin and parity are determined to be~\cite{besjp}  $J^P=1^+$.

The presence of  $Z_c^+(3900)$ in   decays of $b$ hadrons is unclear.
It is not  seen by Belle~\cite{belle2} in the
decay  $\bar B^0 \rightarrow (J/\psi \pi^+) K^-$ nor by LHCb~\cite{lhcbb0} in the decay
 $B^0 \rightarrow (J/\psi \pi^+) \pi^-$.
On the other hand, the $Y(4260)$ may have been seen in the decays
 $B \rightarrow  J/\psi \pi \pi K$ by BaBar~\cite{babar06}, so
  there could  be  production  of $Z_c^+(3900)$  in $b$-hadron  decays
through  the two-step process
 $H_b \rightarrow Y(4260) +$ anything, $Y(4260) \rightarrow Z_c^+(3900) \pi^-$,
 where $H_b$ represents
any hadron containing a $b$ quark.
The process   may be spread over many channels and thus escape
observation in any specific  channel.

In this article   we look for the presence of  such two-step processes 
using  $10.4~\rm{fb^{-1}}$ of $p \overline p $ collision
data  collected by the  D0 experiment at  the Fermilab Tevatron collider.

\section{D0 detector,  event reconstruction  and selection}

The D0 detector~\cite{d0det} has a central tracking system consisting of a silicon
microstrip tracker~\cite{layer0}  and a central scintillating fiber tracker, both located within a
1.9T superconducting solenoidal magnet. A muon
system~\cite{run2muon}  covering pseudorapidity  $|\eta_{\rm det}|<2$~\cite{eta} is 
located  outside of the central tracking system and the liquid argon calorimeter, and
consists of a layer of tracking
detectors and scintillation trigger counters in front of 1.8T toroidal
magnets, followed by two similar layers after the
toroids.

In high-energy $p \overline p$ collisions the $J/\psi$ can be  produced both 
promptly, either directly or in strong interaction decays of higher-mass charmonium states,
or non-promptly in  weak-interaction $b$-hadron decays~\cite{ua1, cdfrun1, d0run1}.
The $b$ and $\bar b$ quarks are produced in pairs and  fragment into the
$b$-hadron species $B^+$, $B^0_d$, $B_s$, $b$ baryons, and $B_c$ with
 the relative branching fractions
0.34, 0.34, 0.10, 0.22, and $<$0.01, respectively~\cite{pdg}.
Non-prompt $J/\psi$ mesons from $H_b$ decays are displaced from the  $p\bar{p}$  interaction vertex
by typically several hundred $\mu$m 
as a result of the long $b$-quark lifetime.

Events used in this analysis are collected with both single-muon and dimuon triggers.
We re-use a  sample of events, prepared for an earlier study of $b$-hadron decays,
containing a non-prompt $J/\psi$ and a pair of oppositely charged particles
consistent with coming from a displaced decay vertex.
For this previously used data sample, the event selection requirement that the 
decay vertex be
separated from the primary vertex with a significance of  more than
$3\sigma$
 precludes extension of the current  study  to include
 the prompt production of $Z_c^+(3900)$ and $Y(4260)$.
Unless indicated otherwise,  we assume the hadrons to be pions and select events in the mass range
 $4.1<m(J/\psi \pi^+ \pi^-)<5.0$~GeV that includes the $Y(4260)$ states and
is high enough for production of the  $Z_c^+(3900)$, but  low enough to exclude 
fully reconstructed direct
decays of $b$ hadrons to final states  $J/\psi h^+ h^-$
where $h$ stands for a pion, a kaon, or  a proton.
In this study of an inclusive final state, we apply more stringent requirements on the 
decay-length-related parameters to further suppress combinations where one of the selected 
particles is produced by the hadronization of partons associated with the primary vertex.

Candidate events  are selected  by requiring  a pair of oppositely charged  muons
and  a charged particle  with $p_T$ above  1~GeV   at a common vertex  with
 $\chi^2 < 10$ for 3 degrees of freedom.   
Muons  must 
   have transverse momentum $p_T > 1.5$~GeV.
At least one muon must traverse both inner and outer layers of the muon detector.
 Both muons must match tracks 
in the central tracking system. The reconstructed invariant mass $m(\mu^+\mu^-)$ 
must be between 2.92 and 3.25~GeV, 
consistent with the world average mass of the $J/\psi$~\cite{pdg}.
To select final states originating from $b$-hadron decays,
the  $J/\psi +1$~track vertex is required to be  displaced from
the $p\bar{p}$  interaction vertex in the transverse plane by at least 5$\sigma$ and
 the transverse impact parameter~\cite{ip}
 significance ${\sl IP}/\sigma({\sl }IP)$ 
of the hadronic track is required to be greater than $2\sigma$.

For accepted $J/\psi +1$~track combinations,  another track,
 with an opposite charge to the first track and  with $p_T>0.8$ GeV,
is added to form  a common $J/\psi +2$~tracks system. 
The second track must have an  ${\sl IP}$ significance greater than 1$\sigma$ 
and its contribution to the  $\chi^2$ of the $J/\psi +2$ tracks   vertex~\cite{vertex} must be  less
than six. The cosine of the  angle in the transverse plane between the momentum vector and decay path
of  the  $J/\psi +2$~tracks system is required to be greater
than 0.9.

\begin{figure}[htb]
\includegraphics[width=0.9\columnwidth]{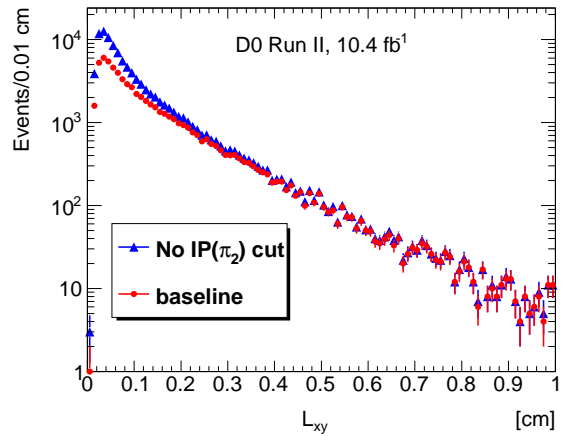}
\caption{\label{fig:Lxy} 
The $J/\psi \pi^+ \pi^-$ decay length  in the transverse plane for accepted candidates in the range
  $4.2<m(J/\psi \pi^+ \pi^-)<4.7$ GeV 
and for
the case when the ${\sl IP}$ cut on the second pion is removed.
}
\end{figure}

For the accepted  $J/\psi +2$~tracks combinations we calculate the $J/\psi \pi^+ \pi^-$
invariant mass  by assigning the pion mass to both hadronic tracks.
We correct the muon momenta by 
 constraining $m(\mu^+\mu^-)$ to the world average $J/\psi$ meson  mass~\cite{pdg}. 
The sample includes events in which  the hadronic pair comes from decays
$K^* \rightarrow K \pi$ or $\phi \rightarrow K K$.
We remove such events by vetoing the mass combinations
$0.81<m(\pi K)<0.97$ GeV, $0.81<m(K \pi)<0.97$ GeV, and  $1.01<m(K K)<1.03$ GeV.
We also veto photon conversions by removing events with  $m(\pi^+ \pi^-)<0.35$ GeV.
The $K^*$ veto rejects about 20\% of the phase space while  reducing  the background
by about a factor of two.  The combination of the three vetoes reduces the background by
a factor  of about 2.5. 
Multiple candidates per event are allowed but their rate is negligible.

The transverse decay length distribution of the $J/\psi \pi^+ \pi^-$ system  $L_{\rm xy}$ is
shown in Fig.~\ref{fig:Lxy}.
With the average resolution of 0.0057~cm most of the prompt events would be
contained at $L_{\rm xy}<0.025$ cm.  The distribution
confirms that prompt background has been strongly suppressed and that  
the  selected  $J/\psi +2$~tracks combinations
originate predominantly from partially reconstructed vertices of  $b$-hadron decays.

\section{Fit results}

\begin{figure*}[htbp]
\includegraphics[width=0.45\textwidth]{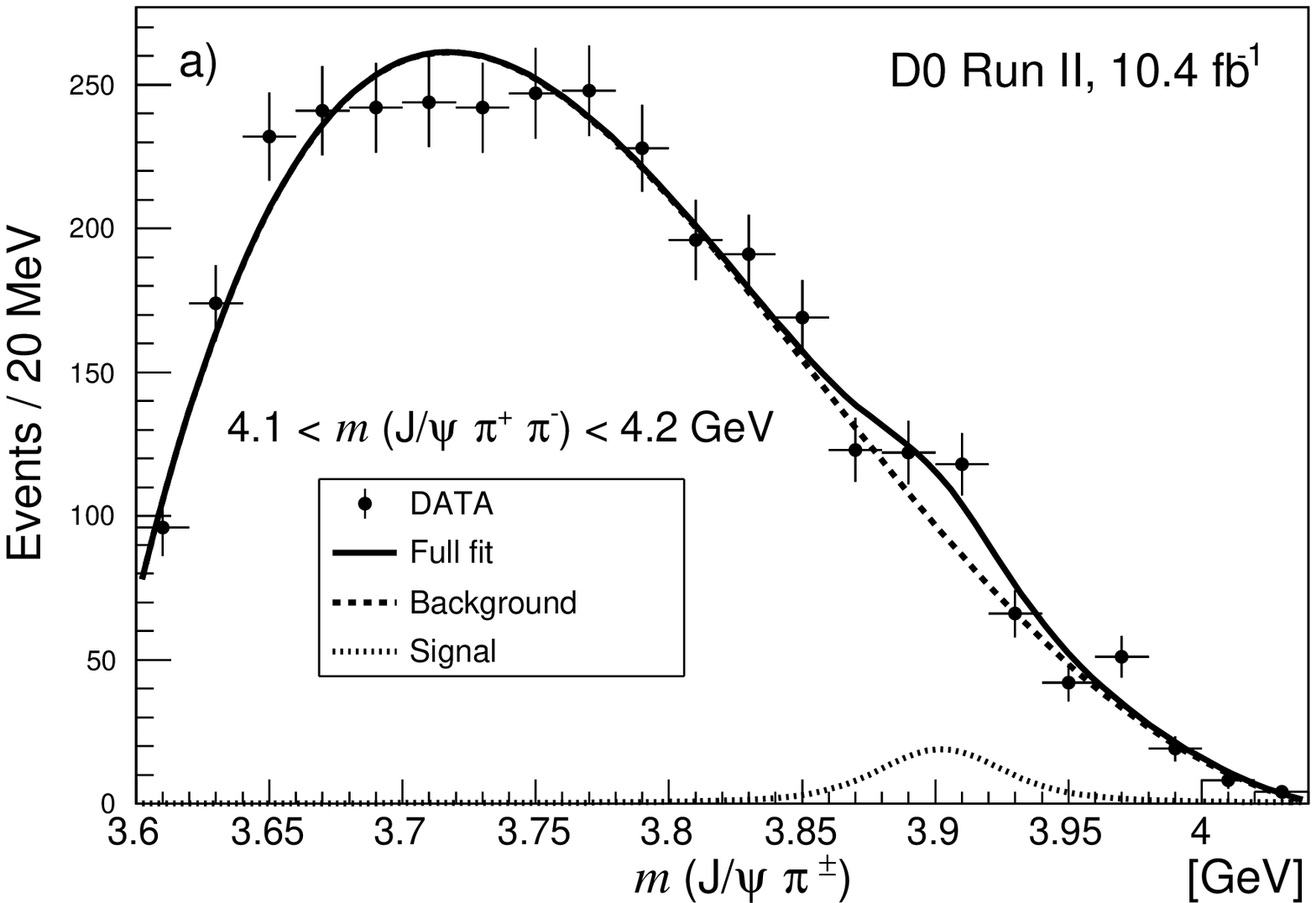}
\includegraphics[width=0.45\textwidth]{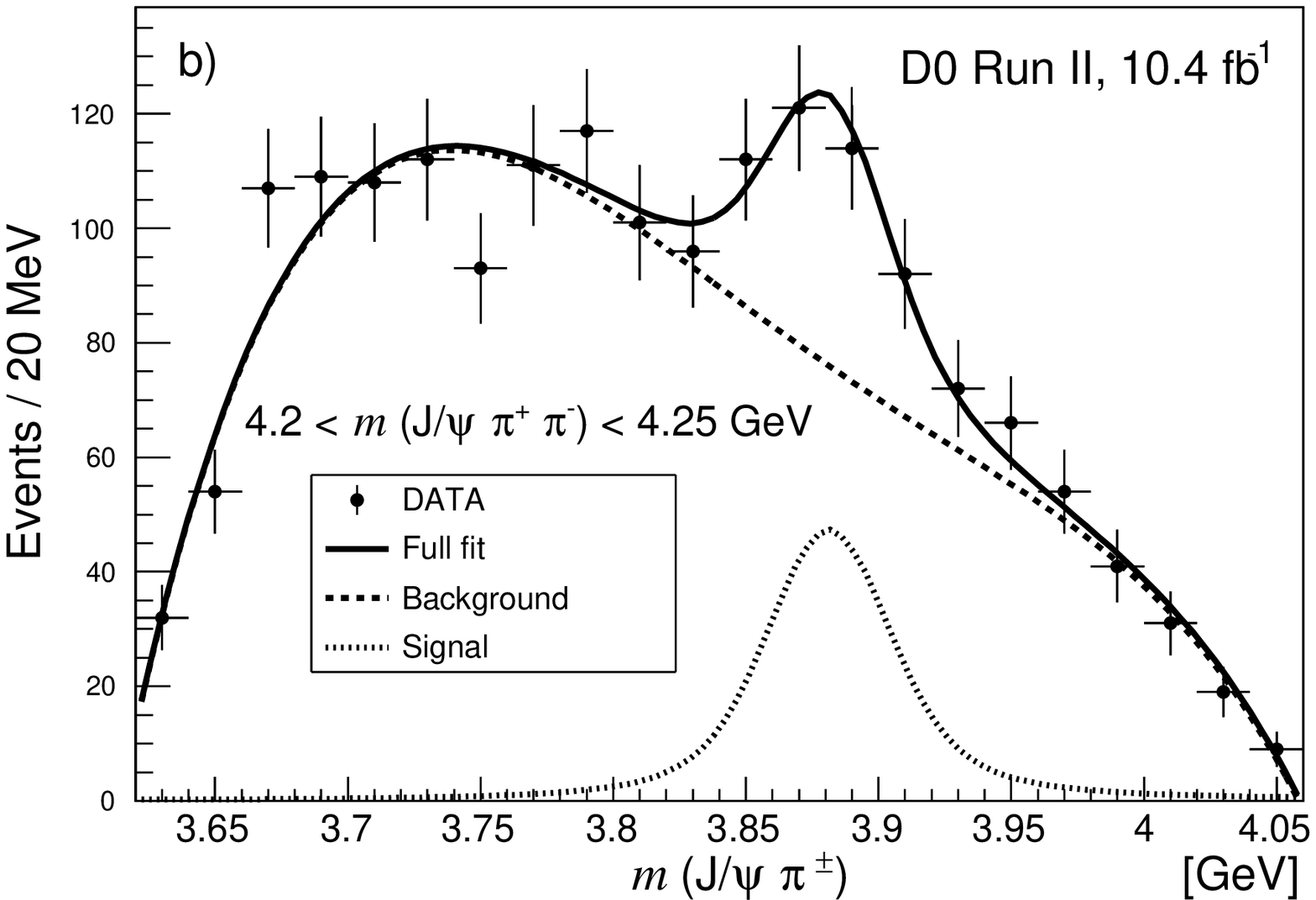}
\includegraphics[width=0.45\textwidth]{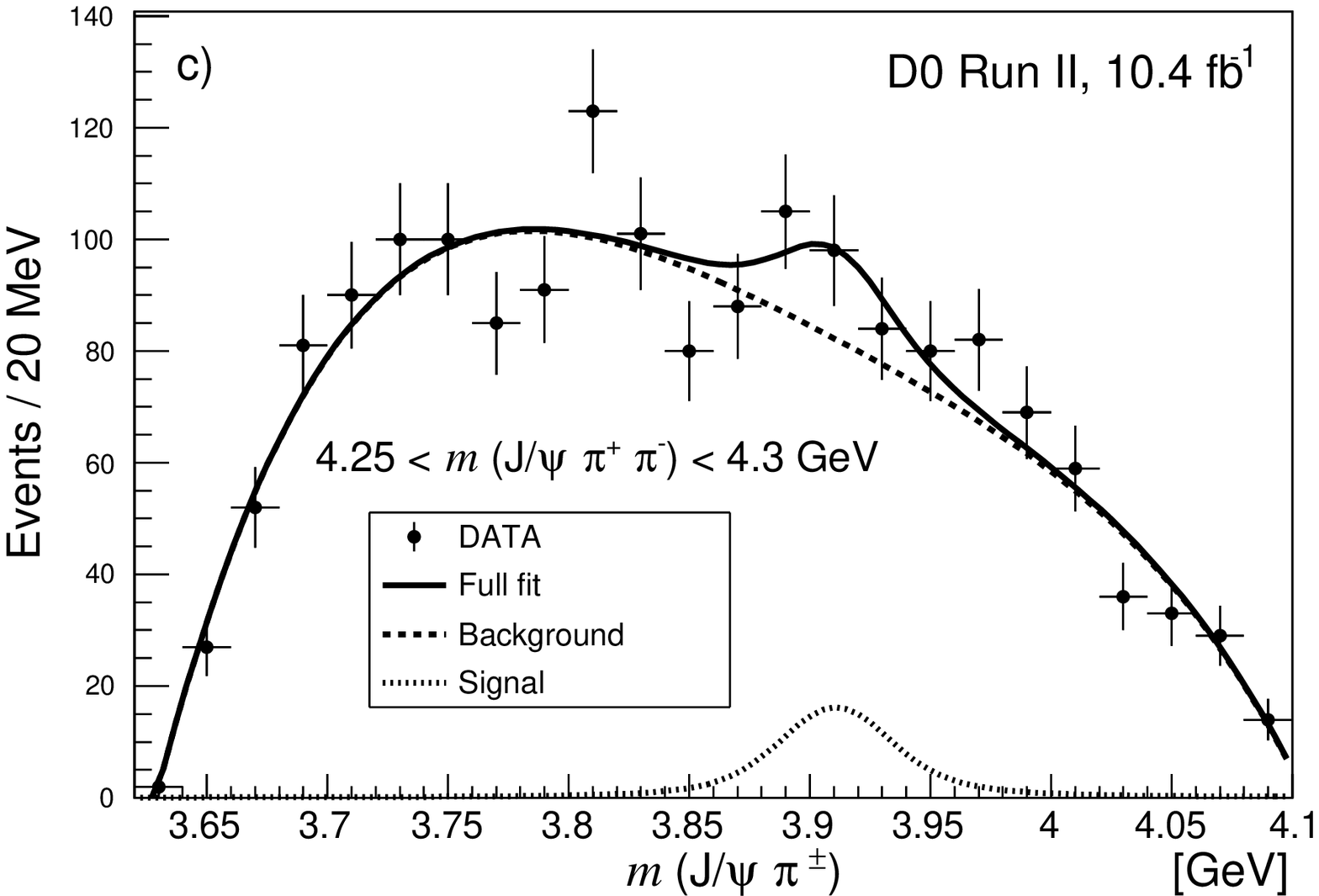}
\includegraphics[width=0.45\textwidth]{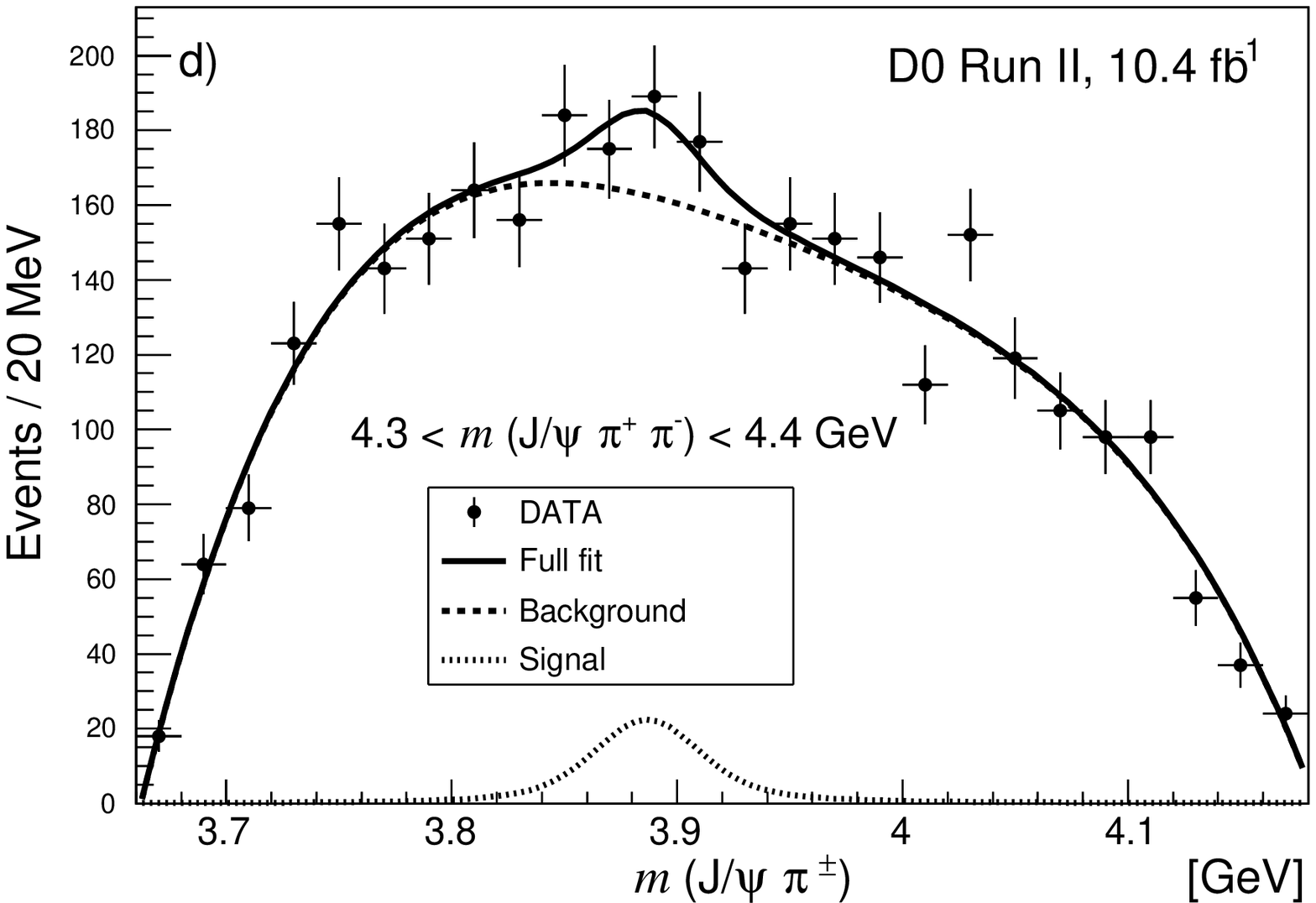}
\includegraphics[width=0.45\textwidth]{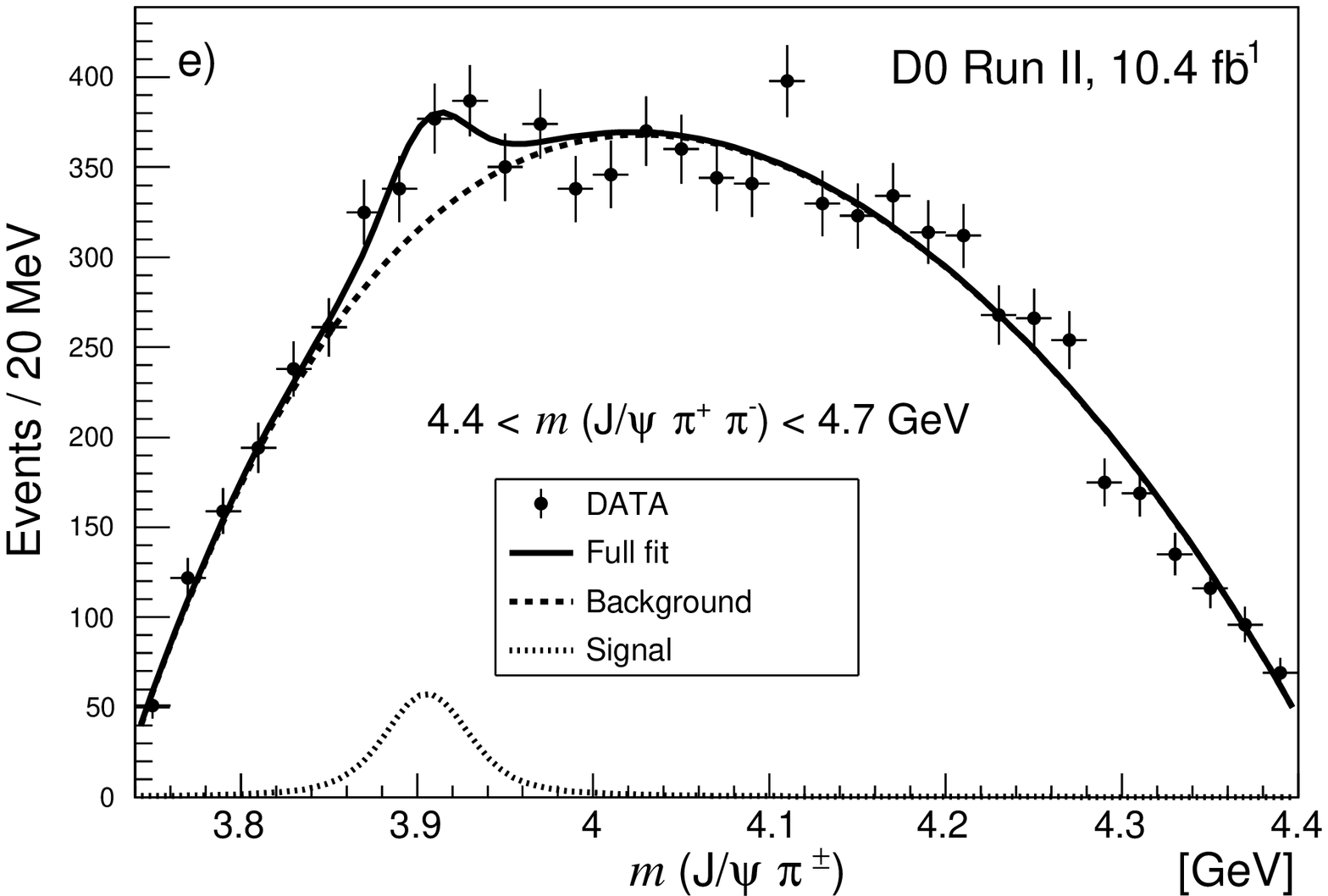}
\includegraphics[width=0.45\textwidth]{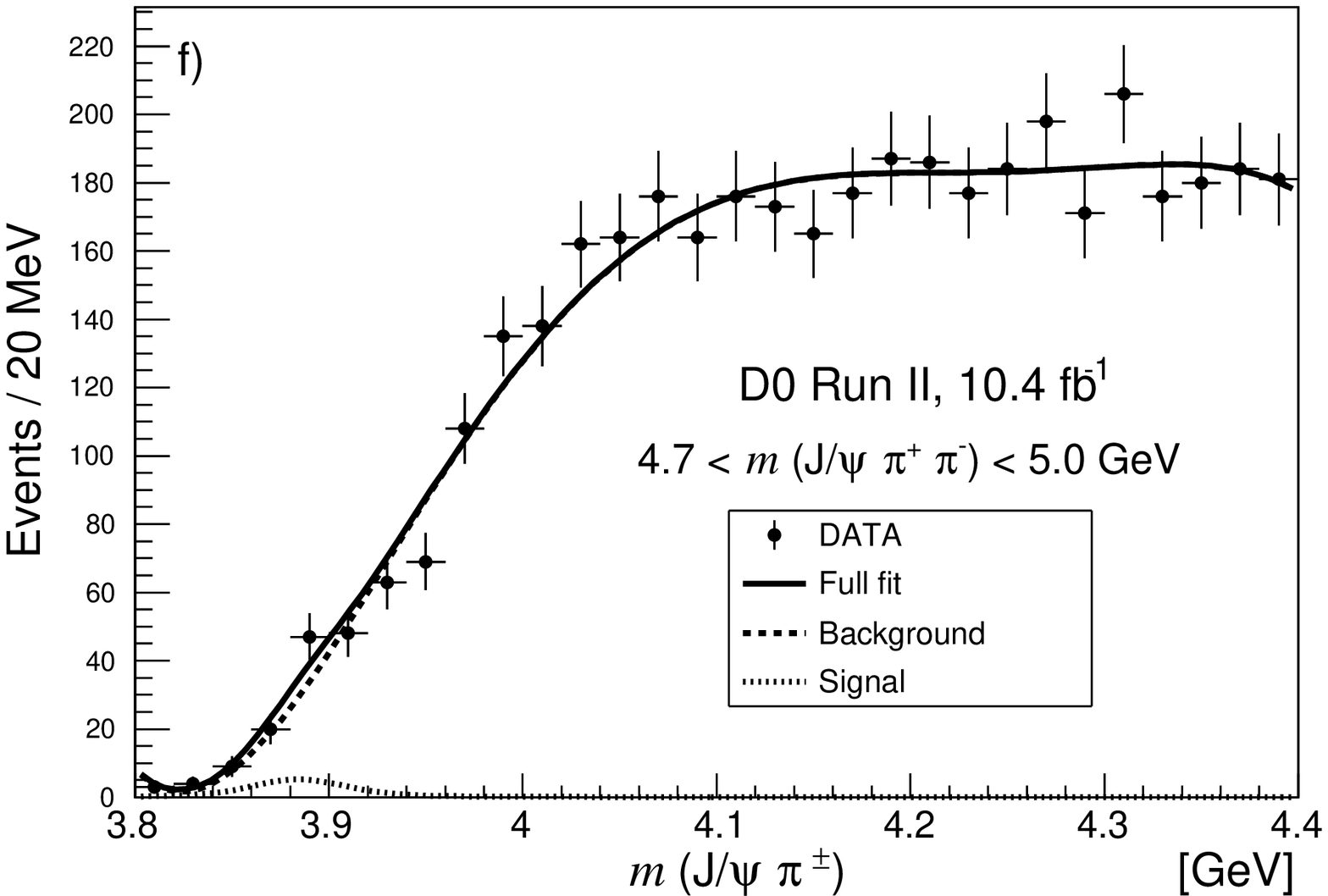}
\caption{\label{fig:z2trky} 
The invariant mass distribution of  $J/\psi \pi^+$ candidates in six ranges  of $m(J/\psi \pi^+ \pi^-)$
as indicated.
The solid lines show the results of the fit. The dashed lines show the combinatorial background
and the dotted lines indicate the signal contributions.
}
\end{figure*}

Our study is focused on  the $J/\psi\pi^+$ system around  the   $Z_c^+(3900)$ mass.
As mentioned above, the production of  $Z_c^+(3900)$ may occur through a
sequential process with an intermediate  $Y(4260)$, e.g., 
 $B^+ \rightarrow Y(4260) K^+$, $Y(4260) \rightarrow Z_c^+(3900) \pi^- $.
To test this possibility, we select events in the mass range   $4.1<m(J/\psi \pi^+ \pi^-)<5.0$~GeV.
We construct the mass  $m(J/\psi\pi^+)$ by combining the $J/\psi$ with either
of the two pion candidates and, following Refs.~\cite{belle2013} and \cite{bes2013},  selecting
 the higher-mass combination.
We fit the resulting  $m(J/\psi\pi^+)$  distribution  to the  sum of a resonant signal  represented by
a  relativistic $S$-wave Breit-Wigner  function
with a  width fixed to  
 $\Gamma=28.2$~MeV~\cite{pdg} smeared with the D0  mass resolution of  $\sigma=17\pm2$~MeV and 
a mass that is allowed to vary freely, 
and an incoherent background.
Background is mainly due to $b$-hadron decays to a  $J/\psi$, with a random
  hadron coming  from the same multi-body decay.
For the background shape we use  Chebyshev polynomials of the first kind.
The fitting range  is  chosen so as to 
obtain an acceptable fit 
while  avoiding  regions where the background function becomes negative.

\begin{figure}[htb]
\includegraphics[width=0.95\columnwidth]{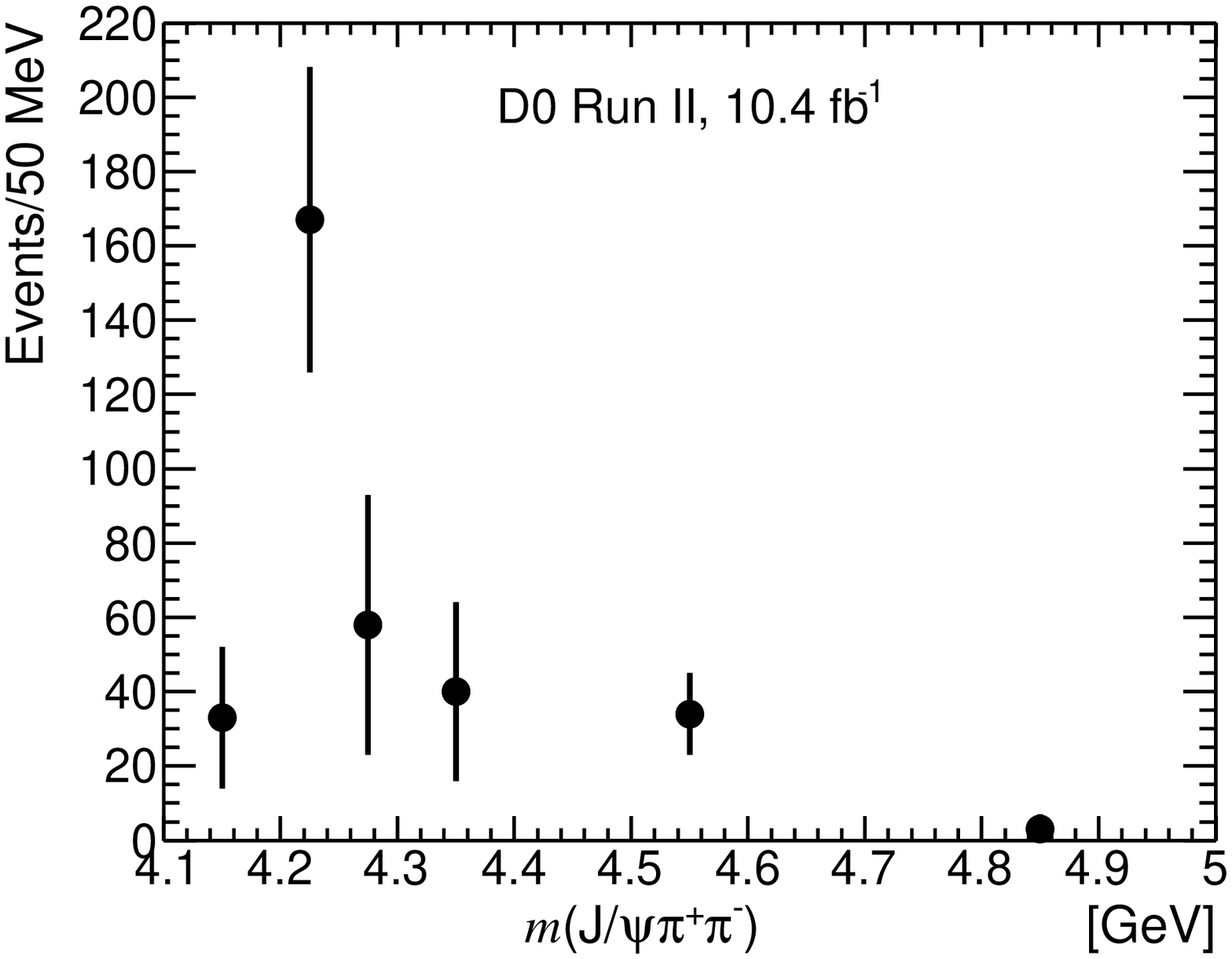}
\caption{\label{fig:zfromy} The  $Z_c^+(3900)$ signal yield per 50 MeV 
 for the six intervals of $m(J/\psi \pi^+ \pi^-)$:
4.1$-$4.2, 4.2$-$4.25, 4.25$-$4.3,
4.3$-$4.4, 4.4$-$4.7, and 4.7$-$5.0~GeV. The points are placed at the bin centers.
}
\end{figure}

\begin{figure}[htbp]

\includegraphics[width=0.95\columnwidth]{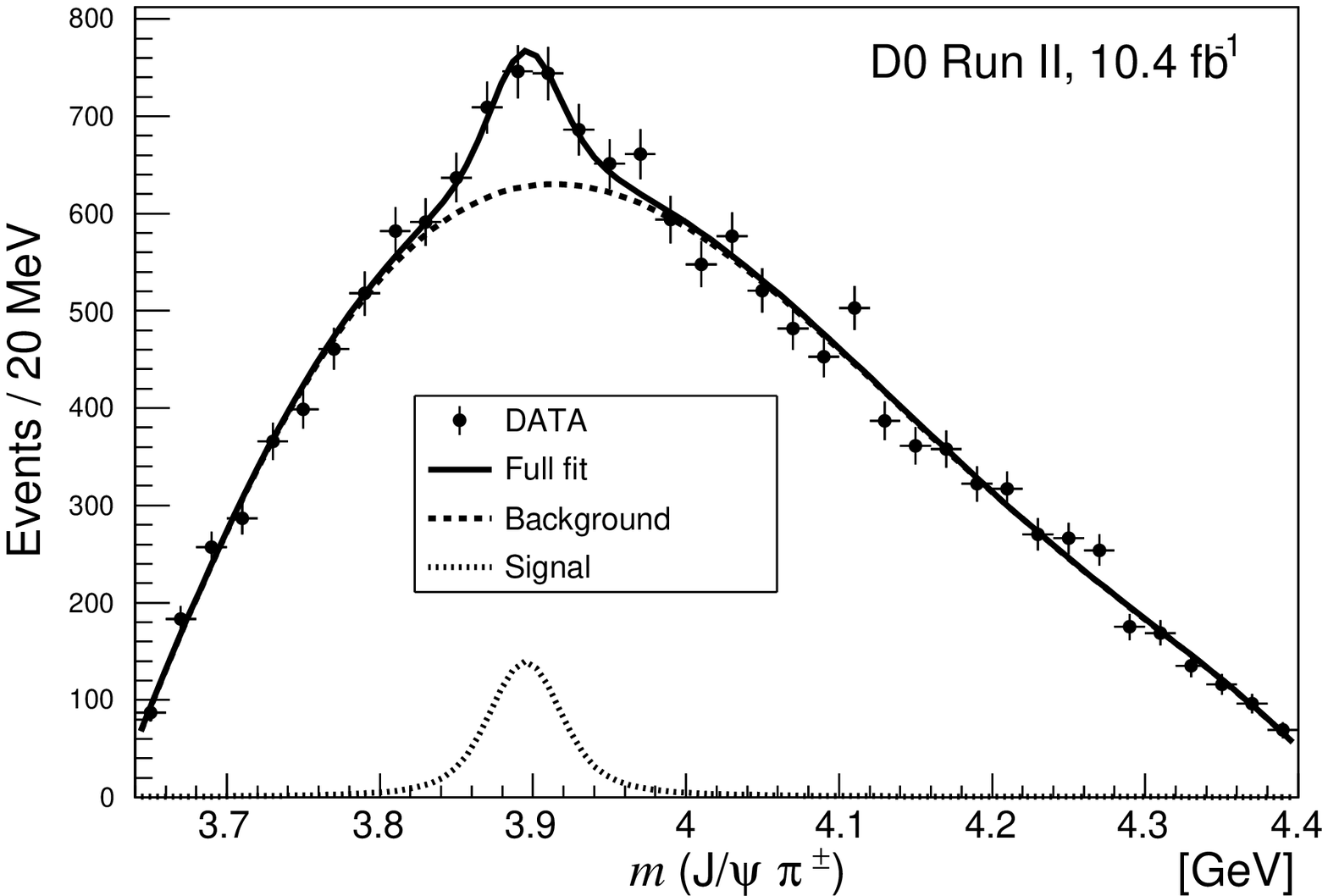}
\caption{\label{fig:z4247} 
The invariant mass distribution of  $J/\psi \pi^+$ candidates in the  range 
$4.2<m(J/\psi \pi^+ \pi^-)<4.7$~GeV.
The solid line shows the result of the fit. The dashed line show the combinatorial background
parametrized with the fifth-order Chebyshev polynomial
and the dotted line indicates the signal contribution.
}
\end{figure}

We   perform  binned maximum-likelihood fits to the  $J/\psi \pi^+$ mass distribution
in six  $J/\psi \pi^+ \pi^-$ mass intervals
of varying size, chosen to align with the  $Y(4260)$  states. 
These intervals, (4.1$-$4.2), (4.2$-$4.25), (4.25$-$4.3), (4.3$-$4.4), (4.4$-$4.7), and
(4.7$-$5.0)~GeV, contain roughly equal numbers of signal plus background events. In each interval we
represent the background contribution by a Chebyshev polynomial whose order is chosen
to  minimize the Aikake Information Criterion  ($AIC$)~\cite{aic}.
For a fit with $p$ free parameters to a distribution in $n$ bins the $AIC$ is defined as
$AIC = \chi^2 +2p + 2p(p+1)/(n-p-1)$. 
We use fourth-order polynomials in all bins except (4.7$-$5.0)~GeV where we use a fifth-order polynomial.

As shown in Fig.~\ref{fig:z2trky}, we see a clear enhancement near  the  $Z_c^+(3900)$ 
mass for events in the range $4.20<m(J/\psi \pi^+ \pi^-)<4.25$~GeV, consistent with coming from
the $\psi(4260)$ (recall that the $\psi(4260)$ mass is 4230 $\pm$~8 MeV~\cite{pdg}), 
  and smaller but finite $Z_C^+(3900)$ signals  for $m(J/\psi \pi^+ \pi^-)$ ranges
between 4.2~GeV and 4.7~GeV.
We find no significant signal in the bins  $4.1<m(J/\psi \pi^+ \pi^-)<4.2$~GeV or
  $4.7<m(J/\psi \pi^+ \pi^-)<5.0$~GeV.
The resulting differential distribution of the signal yield is shown in
Fig.~\ref{fig:zfromy}.
We note the presence of a $Z_c^+(3900)$  signal with a statistical significance 
greater than $3\sigma$ in the $4.4<m(J/\psi \pi^+ \pi^-)<4.7$ GeV region  above 
 the $\psi(4360)$  signal~\cite{bes3y}, indicating  some contribution from
 a non-$Y(4260)$  $J/\psi \pi^+ \pi^-$ combination.
The measured signal masses are consistent with each other (with a p-value of 0.1).

 We then perform a  fit to the data in the mass
 range  $4.2<m(J/\psi \pi^+ \pi^-)<4.7$~GeV.
The $AIC$ test gives similar results using  the fifth- and fourth-order polynomial
as background while the $\chi^2$ test prefers the fifth-order polynomial
(p-value of 0.18 vs. 0.066). 
The fit using the fifth-order polynomial background shown in  Fig.~\ref{fig:z4247}  yields  
 $N=502\pm 92~({\rm stat})$ signal events,
$m=3895.0\pm5.2~({\rm stat}) $ MeV, and a statistical significance of $S=5.6\sigma$.
The fit using the fourth-order polynomial gives 
$N=608\pm82$, $m=3895.7\pm4.6$~MeV,  and $S=7.7\sigma$.
The statistical significance of the signal 
is defined as $S=\sqrt{-2\, {\rm ln} ({\cal{L}}_0 /{\cal{L}}_{\rm max}) }$,
where ${\cal{L}}_{\rm max}$ and ${\cal{L}}_0$  are likelihood values for the
best-fit signal yield and for the signal yield fixed to zero.
In the following we choose the fit using the fifth-order polynomial as the baseline.
A $\chi^2$ test of the fit quality gives the  $\chi^2$ over the number of degrees
of freedom ($\rm ndf$)   $\chi^2/{\rm ndf}=36.8/30$.

\section{Cross-checks}

In an alternative approach, we perform a simultaneous fit to the four  subsamples
of the $m(J/\psi \pi^+ \pi^-)$ in the  4.2$-$4.7~GeV range,
allowing for separate free   parameters  of the fourth-order Chebyshev polynomial background  and
free  signal yields but using a common free signal mass
parameter.  
The fitted mass is 
$3889.6 \pm 9.8$~MeV, and the number of signal events is $444 \pm 149$, in agreement with the baseline
result,
and the quality of the fit is $\chi^2/{\rm ndf}=53.3/81$.

We divide the sample into two ranges of the  $p_T$ of the pion from the $Z_c^+(3900)$ decay,
  $p_T(\pi)<1.5$~GeV and  $p_T(\pi)>1.5$~GeV,
 and fit them
separately.
The fitted yields are  $202\pm51$ and $319\pm72$ events
and the masses are $3906.6\pm10.0$~MeV and $3896.1\pm6.7$~MeV, respectively.

Fits to the three $Z_c^+(3900)$ pseudorapidity  ranges $|\eta|<0.9$, $0.9<|\eta|<1.3$ and $1.3<|\eta|<2.0$
containing similar numbers of events
 give the signal yields of
$195\pm 57$, $155\pm52$, and $163\pm48$ and  mass values of $3902.8\pm7.3$~MeV,
$3906.4\pm11.2$, and $3887.8\pm8.8$~MeV. 
The signal to  background ratios  in the three $|\eta |$ regions are consistent with
 being the same, as would be expected if both signal and the dominant backgrounds
 arise from the decays of $b$ hadrons.

To test the sensitivity of the results to the fit quality requirements, we define a control
sample by selecting events with the  fit quality of the $J/\psi + 1$ track vertex
in the range $10<\chi^2 < 20$. 
The fitted yield in the control sample is $10\pm25$ events, consistent with no signal.

Due to the limited muon momentum resolution, our selection of the $J/\psi$ mass window
passes some non-$J/\psi$ dimuons while rejecting a fraction of genuine $J/\psi$'s.
The  non-$J/\psi$ background includes sequential decays $b \rightarrow c \mu X$,
 $c \rightarrow s \mu X$, and semileptonic $b$-hadron decays accompanied by
a muon track originating from a charged pion or kaon decay in flight.
We estimate the fraction of non-$J/\psi$ background in our baseline sample 
at 9\% and the dimuon mass cut efficiency for $J/\psi$ at  94\%.
A fit  to the $m(J/\psi \pi^+)$ spectrum when  the   $J/\psi$ mass window is expanded
  to 2.8--3.4~GeV
yields  $530\pm100$ $Z_c^+(3900)$ signal events, 6\% more than in the baseline analysis,
 in agreement with expectation.

\section{Systematic uncertainties}

There are several sources of systematic uncertainties in the  baseline measurement of the  $Z_c^+(3900)$ 
mass and yield, summarized in Table~\ref{tab:syst}.

\begin{table}[h]
\caption{\label{tab:syst} Systematic uncertainties for the  $Z_c^+(3900)$
mass and yield measurements. }
\begin{ruledtabular}
\def\arraystretch{1.1}
\begin{tabular}{lccc}
Systematic uncertainty & Mass (MeV)   & Yield\\
\hline
Mass calibration & $^{+3}_{-0}$ & $<$1\\
Mass resolution   & $<0.1$  &  $\pm 27$ \\
Background shape & $\pm0.4$ & $\pm53$   \\
Bin size   & $\pm 1.1$  &  $\pm 9$\\
Signal model & $\pm 2.4$ & $\pm3$\\
Natural width variation & $<$0.1 &  $\pm 23$\\
\hline 
Total (sum in quadrature) & $-2.7,+4.0$& $\pm 64$\\
\end{tabular}
\end{ruledtabular}
\end{table}

We   assign an asymmetric uncertainty of $(+3,-0)$~MeV  to the $J/\psi \pi^+$ mass scale
based on studies of  the D0 measured mass shift  compared to world-average values
in several final states with a similar topology~\cite{incl}.

The estimate of the mass resolution  is based on
the dependence of the measured and simulated resolution of
the released kinetic energy  for decays with a similar topology.
The variation of the assumed resolution by its uncertainty of $\pm 2$~MeV
has a negligible effect on the measured  $Z_c^+(3900)$ mass.
We assign an uncertainty on  the  signal yield
equal to half of the difference between the two extreme results.

We assess the effects of the fitting procedure and background  shape
as half of the difference of the results obtained with the fourth- and fifth-order
Chebyshev polynomial. Similarly, we estimate  the effect of bin size by
comparing the results for 20~MeV and 10~MeV bins.

We assign the uncertainty in the signal model as  half of the difference
in the   results obtained with  the  relativistic   Breit-Wigner shapes
with  and without the energy dependence of the natural width.

In the analysis we set the natural width equal to the world-average value.
We assign the uncertainty in the mass and yield measurement
by repeating the fits with the width altered by $\pm 2.6$~MeV~\cite{pdg}.

\section{Results}

\subsection{The {\boldmath $Z_c(3900)$} signal yield as a function of  {\boldmath $m(J/\psi \pi^+ \pi^-)$}}

Table~\ref{tab:results} lists the  $Z_c^+(3900)$ fitted signal yields and the measured mass
in the six non-overlapping intervals of the $J/\psi \pi^+ \pi^-$ invariant mass between 4.1~GeV and
5.0~GeV. The $Z_c^+(3900)$ width is fixed at $\Gamma=28.2$~MeV for these fits.
The measured masses are consistent with each other and with the 
original results of Refs.~\cite{belle2013} and \cite{bes2013}, and thus
we conclude that we are observing the same $Z_c^+(3900)$ state. We report the results for the 
range 4.2$-$4.7~GeV  as our best measurement of the mass
 of the  $Z_c^+(3900)$ resonance and the signal significance.

Our baseline result above allows the $Z_c^+(3900)$ mass to float but fixes its width at the world average value, and thus raises the question of whether the significance of the fit would change if the world average [4] mass were used.  We have tested this by fixing the mass to $m = 3886.6$ MeV [4].  The fit gives a yield
of $480 \pm 91$, $\chi^2/{\rm ndf} = 39/31$, and significance $S=5.4\sigma$ that differ very little 
from our baseline result. A slightly better fit is obtained 
 with the mass and width fixed to the PDG values~[4] for just those measurements that use the final state $Z_c^{\pm,0} \rightarrow J/\psi \pi^{\pm,0}$:  $m = 3893.3$ MeV and $\Gamma = 36.8$ MeV.  In this case we obtain $\chi^2$/ndf = 35.9/31, yield 
of $580 \pm 104$ and  $S=5.7\sigma$.  We conclude that variations in the choice of $Z_c^+$ mass and width have only a small effect upon our conclusions.

 \begin{table}[h]
\begin{center}
\caption{ $Z_c^+(3900)$ signal yields and mass measurements, fit quality, and statistical significance $S$ 
in intervals of $m(J/\psi \pi^+ \pi^-)$.
The  six measurements in non-overlapping subsamples are dominated by statistical
uncertainties. There is a common asymmetric $+3,-0$~MeV mass uncertainty.
The last row shows a summary result that includes  statistical and systematic uncertainties. }
\def\arraystretch{1.0}
\begin{tabular}{ccccc}\hline \hline 
$m(J/\psi \pi^+ \pi^-)$   & Event yield &  Mass  & $\chi^2/{\rm ndf}$ & $S$  \\
(GeV)  & & (MeV)  & &  ($\sigma$) \\ \hline
4.1$-$4.2           & $66\pm 38$ & $3902.2\pm 10.6$  & 24.1/15 & 1.7  \\    
4.2$-$4.25           & $167\pm 41 $ & $3881.3\pm 6.1$ & 14.6/15 & 4.3   \\    
4.25$-$4.3           & $58\pm 35  $ & $3910.7\pm 15.7$ &  23.6/17 & 1.6 \\  
4.3$-$4.4           & $80\pm 48  $ & $3886.5\pm 13.0$ & 26.3/19 & 1.8  \\
4.4$-$4.7           & $206\pm 65 $ & $3905.7\pm 9.5$ & 35.8/26 & 3.2  \\ 
4.7$-$5.0           & $19\pm 25 $ & $3884.7\pm 26.6$ & 21/22 & 0.4  \\ \hline
4.2$-$4.7           & $502\pm 92 \pm 64$ & $3895.0\pm 5.2 ^{+4.0}_{-2.7} $  & 36.8/30 & 4.6 \\    
\hline \hline
\end{tabular}
\label{tab:results}
\end{center}
\end{table}

 The systematic uncertainties are 
taken into account in the estimate of the significance 
by convolving the p-value as a function of signal yield
with a Gaussian function with a mean corresponding to our measured value and 
width equal to the systematic uncertainty on the yield.
Adding the systematic uncertainty changes the significance for the baseline
fit from 5.6$\sigma$ to 4.6$\sigma$.

\subsection{Normalization to {\boldmath $B_d^0 \rightarrow J/\psi K^*$}}

\begin{figure}[htb]
\includegraphics[width=0.95\columnwidth]{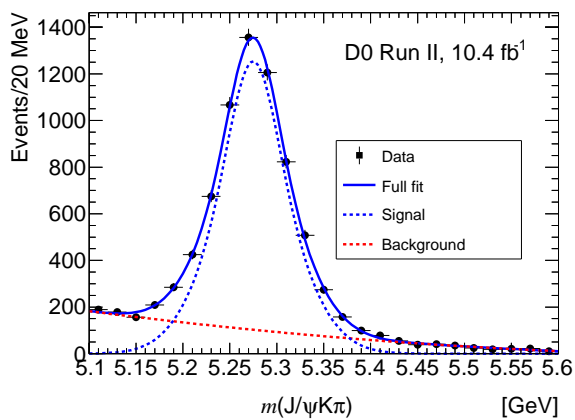}
\caption{\label{fig:mb0} 
The invariant mass distribution  of  accepted  $J/\psi +2$ track candidates under
 the  $J/\psi K^{\pm}\pi^{\mp}$
hypothesis with a requirement that (at least) one of the $K^{\pm}\pi^{\mp}$ combinations is within the $K^*$ window
(see text). 
}
\end{figure}

We normalize the  $Z_c^+(3900) \rightarrow J/\psi \pi^+$ signal in the parent $J/\psi \pi^+ \pi^-$ mass
range of 4.2$-$4.7~GeV to the number of events of the decay $B_d^0 \rightarrow J/\psi K^*$.
The latter are required to satisfy the same stringent kinematic and quality cuts as applied
to the  $J/\psi \pi^+ \pi^-$ except that the $K^*$ veto is replaced with the requirement
that at least one $K^{\pm}\pi^{\mp}$ pair is within the $K^*$ mass window. If two such pairs are present
 we select the $K^{\pm}\pi^{\mp}$ combination 
with mass closer to the $K^*$ mass. 
We fit the distribution to a sum of a signal described by a double Gaussian function
and a quadratic polynomial background.
We find  the  number of $B^0_d$ decays
$N(B^0_d)= 5900 \pm 116~({\rm stat})$  and  obtain the ratio of 
the observed number of events 
$502/5900=0.085\pm0.019$ where the uncertainty 
is a sum in quadrature of the statistical and systematic uncertainties (0.016 and 0.011, respectively).
Since the two processes have the same topology and the kinematic restrictions assure
a uniform track finding efficiency, we assume that the efficiency factors cancel out in the ratio.
The invariant mass $J/\psi K\pi$ distribution and the fit results are
shown in  Fig.~\ref{fig:mb0}.

\begin{figure}[htb]
\includegraphics[scale=0.4]{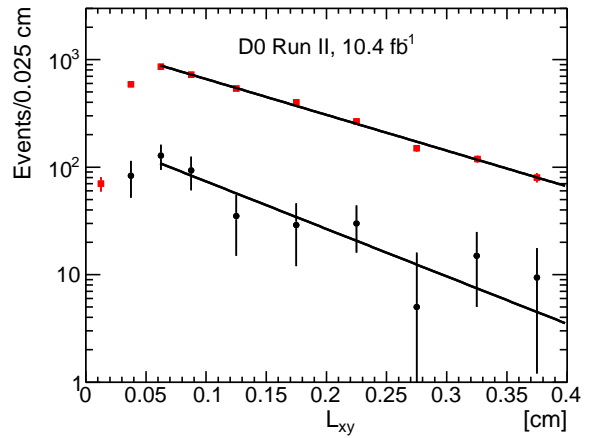}
\caption{\label{fig:lxy} 
The decay length distribution of    $Z_c^+(3900)$ events (black  circles)
and 
$B^0_d \rightarrow J/\psi  K^*$ events (red squares).
}
\end{figure}

\begin{figure}[htb]
\includegraphics[scale=0.4]{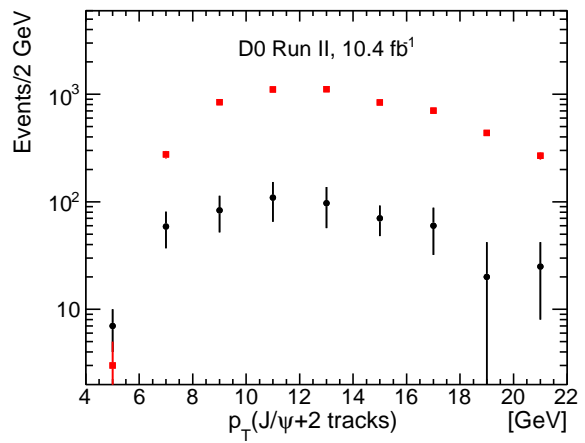}
\caption{\label{fig:pt} 
The $p_T$  of  the $J/\psi \pi^+ \pi^-$ parents of the  $Z_c^+(3900)$ events  (black  circles)
 and  of the $B^0_d$ in the 
$J/\psi  K^*$ channel  (red squares).
}
\end{figure}

Figure \ref{fig:lxy} shows a comparison of    the decay length  distribution of the $Z_c^+(3900)$
 signal events, obtained
by fitting $m(J/\psi \pi^+)$ in bins of the decay length,
and that of  the $B^0_d$ signal from the
 $B^0_d \rightarrow J/\psi K^*$ decay. The mean lifetime of a $b$-hadron admixture averaged
over all $b$ species is similar to the $B^0_d$ lifetime, and the momentum distributions
are similar.
We therefore  expect the  decay length distribution of the two states to show general agreement.
The  distributions show exponential behavior
$N \sim e^{-L_{\rm xy}/\Lambda}$
 in the region above $L_{\rm xy}=0.025$~cm
where the efficiency is constant,
 with consistent coefficients of 
$\Lambda=0.098\pm0.030$ and $0.130\pm0.004$ cm for the $Z_c^+(3900)$ and $B^0_d$, respectively,
  supporting the claim
that the signal events come from  $b$-hadron decays.
The turnover at low $L_{\rm xy}$  occurs because  some  events whose $L_{\rm xy}$ resolution is small
can pass the 5$\sigma$ significance cut for lower values  of   $L_{\rm xy}$.
Figure \ref{fig:pt} compares   the $p_T$   distribution of the 
$J/\psi \pi^+ \pi^-$  system in  $Z_c^+(3900)$  events and the $p_T$ distribution
of $B^0_d$ in the $J/\psi K^*$ channel. The two distributions are similar, as expected
for decay products of $b$ hadrons.
The average $p_T$ of the former (12.5~GeV) 
is lower than the average $p_T$ of  $B^0_d$ (13.6~GeV)
because the $J/\psi \pi^+ \pi^-$ system  carries less than 100\% of the parent $b$ hadron's momentum. 

\subsection{Search for the  {\boldmath $Z_c^+(3900)$} in the decay {\boldmath $\bar B^0_d \rightarrow J/\psi \pi^+ K^-$}}

\begin{figure}[htb]
\includegraphics[scale=0.4]{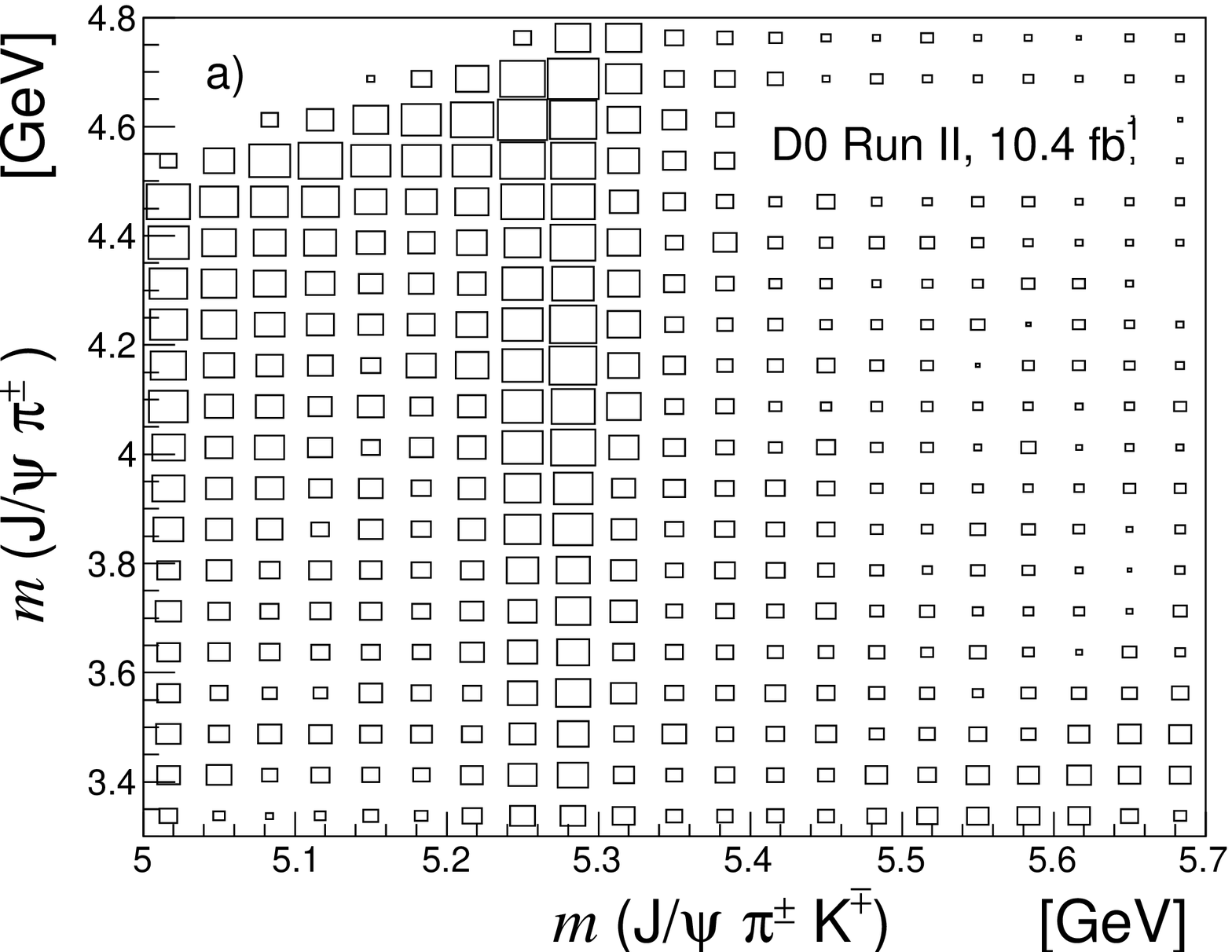}
\includegraphics[scale=0.38]{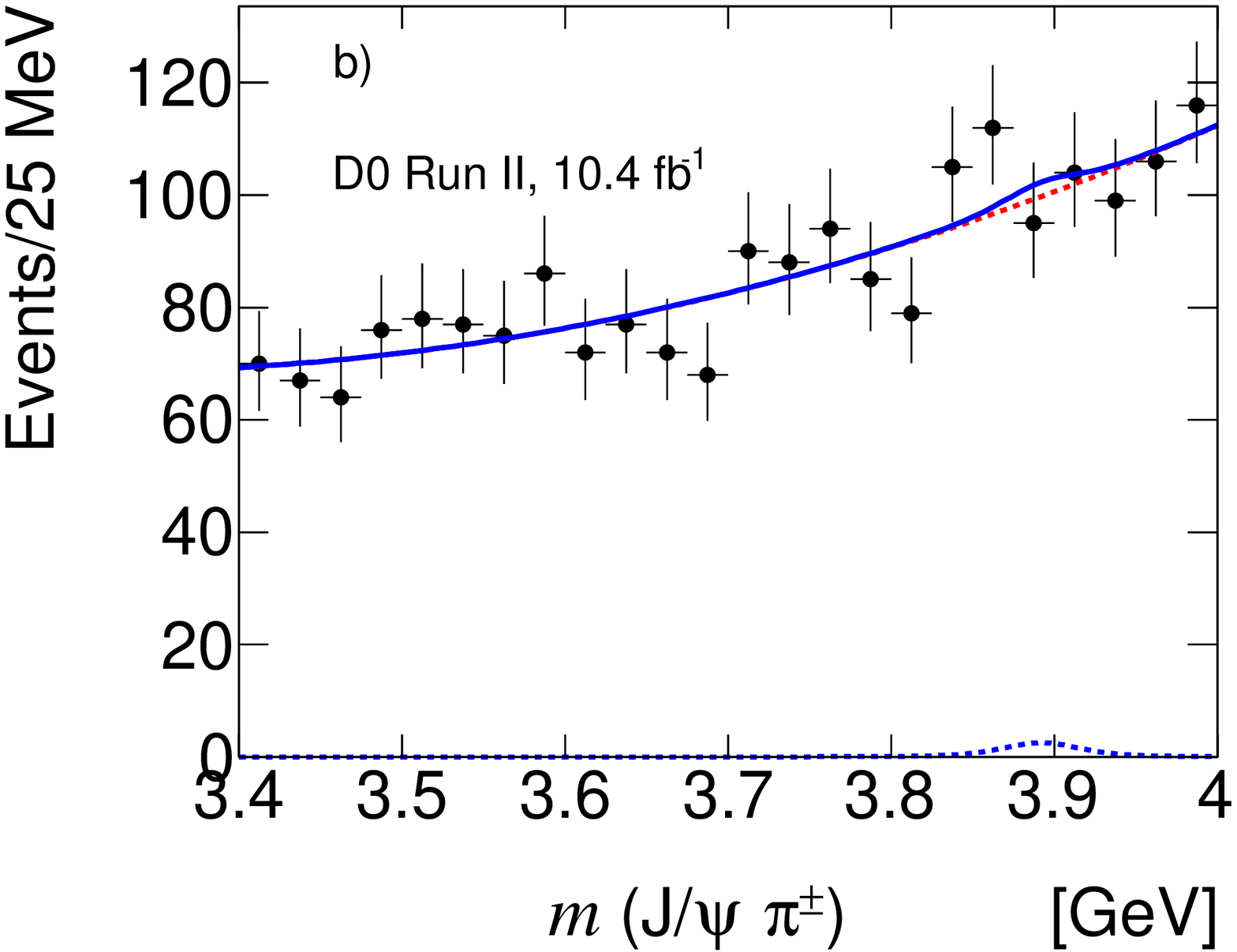}
\caption{\label{fig:zfromb0} 
(a) The scatter plot of $m(J/\psi \pi^+)$ vs.  $m(J/\psi \pi^+ K^-)$
  in the decay  $\bar B^0_d \rightarrow J/\psi \pi^+ K^-$ with the $K^*$ mass range removed.
(b)  The  $m(J/\psi \pi^+)$ distribution in a limited range, for events in the $B^0_d$ mass
window defined as  $5.15<m(J/\psi \pi^+ K^-)<5.4$~GeV,
and a fit allowing for a  $Z_c^+(3900)$ signal and a quadratic background.
}
\end{figure}

 As mentioned in Section I, the Belle Collaboration~\cite{belle2} did not see a significant signal
of the  $Z_c^+(3900)$ in the decay $\bar B^0 \rightarrow J/\psi \pi^+ K^-$. Their amplitude
analysis confirmed the $Z_c(4430)$ and led to an observation of a new resonance, $Z_c(4200)$.
We have studied the $J/\psi \pi^+$ mass in events consistent with this decay,
excluding the events consistent with the decay $\bar B^0_d \rightarrow J/\psi K^*$. 
Figure ~\ref{fig:zfromb0}(a) shows the scatter plot of $m(J/\psi \pi^+)$ vs.  $m(J/\psi \pi^+ K^-)$. 
There is no indication of the   $Z_c^+(3900)$ and the spectrum of   $m(J/\psi \pi^+)$ above 4~GeV
is consistent with the resonance structures observed in Fig. 8 of  Ref.~\cite{belle2}.
Figure~\ref{fig:zfromb0}(b) shows the  $m(J/\psi \pi^+)$ distribution in a limited range
and a fit allowing for a  $Z_c^+(3900)$ signal and a quadratic background.
The fit  gives  an upper limit of 90 signal events at 90\% C.L.
Normalizing to the  5900 events of the $B_d^0 \rightarrow J/\psi K^*$ decay,
we obtain an upper limit on the ratio of the two processes of 0.015,
to be compared to  a limit of 0.0011 obtained by Belle.

\section{Summary and conclusions}

In summary, our study of the semi-inclusive decays of $b$ hadrons 
 $H_b \rightarrow J/\psi  \pi^+  \pi^-$ + anything  reveals a $Z_c^{\pm}(3900)$
signal  that  is correlated with the  $J/\psi \pi^+ \pi^-$ system
in the invariant mass range 4.2$-$4.7~GeV that would  include 
the neutral charmonium-like states $\psi(4260)$ and $\psi(4360)$~\cite{pdg}.
There is an indication that some events arise from $H_b$  decays to an intermediate
 $J/\psi \pi^+ \pi^-$  combination with mass above that of the $\psi(4360)$, with subsequent decay to
$Z_c^{\pm}(3900) \pi^{\mp}$.

The measured mass of the  $Z_c^{\pm}(3900)$ resonance is
$m=3895.0\pm5.2 {\rm \thinspace (stat)} ^{+4.0}_{-2.7}{\rm \thinspace (syst)}$~MeV.
The significance, including  systematic uncertainties, is 4.6 standard deviations. 
We confirm the conclusion of  Ref.~\cite{belle2} that there is no significant
production of the $Z_c^+(3900)$ in the decay $\bar B^0_d \rightarrow J/\psi \pi^+ K^-$.
We set an upper  limit on  the rate of the process $B^0_d \rightarrow Z_c^+(3900) K^-$ 
relative to  $B_d^0 \rightarrow J/\psi K^*$ at 0.015 at the 90\% C.L.
With the present data sample  we have no sensitivity to  prompt production of the  $Z_c^{\pm}(3900)$ 
in $p \overline p$ collisions.

\input acknowledgement_APS_full_names.tex
\end{document}

%% file: author_list.tex
\affiliation{LAFEX, Centro Brasileiro de Pesquisas F\'{i}sicas, Rio de Janeiro, RJ 22290, Brazil}
\affiliation{Universidade do Estado do Rio de Janeiro, Rio de Janeiro, RJ 20550, Brazil}
\affiliation{Universidade Federal do ABC, Santo Andr\'e, SP 09210, Brazil}
\affiliation{University of Science and Technology of China, Hefei 230026, People's Republic of China}
\affiliation{Universidad de los Andes, Bogot\'a, 111711, Colombia}
\affiliation{Charles University, Faculty of Mathematics and Physics, Center for Particle Physics, 116 36 Prague 1, Czech Republic}
\affiliation{Czech Technical University in Prague, 116 36 Prague 6, Czech Republic}
\affiliation{Institute of Physics, Academy of Sciences of the Czech Republic, 182 21 Prague, Czech Republic}
\affiliation{Universidad San Francisco de Quito, Quito 170157, Ecuador}
\affiliation{LPC, Universit\'e Blaise Pascal, CNRS/IN2P3, Clermont, F-63178 Aubi\`ere Cedex, France}
\affiliation{LPSC, Universit\'e Joseph Fourier Grenoble 1, CNRS/IN2P3, Institut National Polytechnique de Grenoble, F-38026 Grenoble Cedex, France}
\affiliation{CPPM, Aix-Marseille Universit\'e, CNRS/IN2P3, F-13288 Marseille Cedex 09, France}
\affiliation{LAL, Univ. Paris-Sud, CNRS/IN2P3, Universit\'e Paris-Saclay, F-91898 Orsay Cedex, France}
\affiliation{LPNHE, Universit\'es Paris VI and VII, CNRS/IN2P3, F-75005 Paris, France}
\affiliation{CEA Saclay, Irfu, SPP, F-91191 Gif-Sur-Yvette Cedex, France}
\affiliation{IPHC, Universit\'e de Strasbourg, CNRS/IN2P3, F-67037 Strasbourg, France}
\affiliation{IPNL, Universit\'e Lyon 1, CNRS/IN2P3, F-69622 Villeurbanne Cedex, France and Universit\'e de Lyon, F-69361 Lyon CEDEX 07, France}
\affiliation{III. Physikalisches Institut A, RWTH Aachen University, 52056 Aachen, Germany}
\affiliation{Physikalisches Institut, Universit\"at Freiburg, 79085 Freiburg, Germany}
\affiliation{II. Physikalisches Institut, Georg-August-Universit\"at G\"ottingen, 37073 G\"ottingen, Germany}
\affiliation{Institut f\"ur Physik, Universit\"at Mainz, 55099 Mainz, Germany}
\affiliation{Ludwig-Maximilians-Universit\"at M\"unchen, 80539 M\"unchen, Germany}
\affiliation{Panjab University, Chandigarh 160014, India}
\affiliation{Delhi University, Delhi-110 007, India}
\affiliation{Tata Institute of Fundamental Research, Mumbai-400 005, India}
\affiliation{University College Dublin, Dublin 4, Ireland}
\affiliation{Korea Detector Laboratory, Korea University, Seoul, 02841, Korea}
\affiliation{CINVESTAV, Mexico City 07360, Mexico}
\affiliation{Nikhef, Science Park, 1098 XG Amsterdam, the Netherlands}
\affiliation{Radboud University Nijmegen, 6525 AJ Nijmegen, the Netherlands}
\affiliation{Joint Institute for Nuclear Research, Dubna 141980, Russia}
\affiliation{Institute for Theoretical and Experimental Physics, Moscow 117259, Russia}
\affiliation{Moscow State University, Moscow 119991, Russia}
\affiliation{Institute for High Energy Physics, Protvino, Moscow region 142281, Russia}
\affiliation{Petersburg Nuclear Physics Institute, St. Petersburg 188300, Russia}
\affiliation{Instituci\'{o} Catalana de Recerca i Estudis Avan\c{c}ats (ICREA) and Institut de F\'{i}sica d'Altes Energies (IFAE), 08193 Bellaterra (Barcelona), Spain}
\affiliation{Uppsala University, 751 05 Uppsala, Sweden}
\affiliation{Taras Shevchenko National University of Kyiv, Kiev, 01601, Ukraine}
\affiliation{Lancaster University, Lancaster LA1 4YB, United Kingdom}
\affiliation{Imperial College London, London SW7 2AZ, United Kingdom}
\affiliation{The University of Manchester, Manchester M13 9PL, United Kingdom}
\affiliation{University of Arizona, Tucson, Arizona 85721, USA}
\affiliation{University of California Riverside, Riverside, California 92521, USA}
\affiliation{Florida State University, Tallahassee, Florida 32306, USA}
\affiliation{Fermi National Accelerator Laboratory, Batavia, Illinois 60510, USA}
\affiliation{University of Illinois at Chicago, Chicago, Illinois 60607, USA}
\affiliation{Northern Illinois University, DeKalb, Illinois 60115, USA}
\affiliation{Northwestern University, Evanston, Illinois 60208, USA}
\affiliation{Indiana University, Bloomington, Indiana 47405, USA}
\affiliation{Purdue University Calumet, Hammond, Indiana 46323, USA}
\affiliation{University of Notre Dame, Notre Dame, Indiana 46556, USA}
\affiliation{Iowa State University, Ames, Iowa 50011, USA}
\affiliation{University of Kansas, Lawrence, Kansas 66045, USA}
\affiliation{Louisiana Tech University, Ruston, Louisiana 71272, USA}
\affiliation{Northeastern University, Boston, Massachusetts 02115, USA}
\affiliation{University of Michigan, Ann Arbor, Michigan 48109, USA}
\affiliation{Michigan State University, East Lansing, Michigan 48824, USA}
\affiliation{University of Mississippi, University, Mississippi 38677, USA}
\affiliation{University of Nebraska, Lincoln, Nebraska 68588, USA}
\affiliation{Rutgers University, Piscataway, New Jersey 08855, USA}
\affiliation{Princeton University, Princeton, New Jersey 08544, USA}
\affiliation{State University of New York, Buffalo, New York 14260, USA}
\affiliation{University of Rochester, Rochester, New York 14627, USA}
\affiliation{State University of New York, Stony Brook, New York 11794, USA}
\affiliation{Brookhaven National Laboratory, Upton, New York 11973, USA}
\affiliation{Langston University, Langston, Oklahoma 73050, USA}
\affiliation{University of Oklahoma, Norman, Oklahoma 73019, USA}
\affiliation{Oklahoma State University, Stillwater, Oklahoma 74078, USA}
\affiliation{Oregon State University, Corvallis, Oregon 97331, USA}
\affiliation{Brown University, Providence, Rhode Island 02912, USA}
\affiliation{University of Texas, Arlington, Texas 76019, USA}
\affiliation{Southern Methodist University, Dallas, Texas 75275, USA}
\affiliation{Rice University, Houston, Texas 77005, USA}
\affiliation{University of Virginia, Charlottesville, Virginia 22904, USA}
\affiliation{University of Washington, Seattle, Washington 98195, USA}
\author{V.M.~Abazov} \affiliation{Joint Institute for Nuclear Research, Dubna 141980, Russia}
\author{B.~Abbott} \affiliation{University of Oklahoma, Norman, Oklahoma 73019, USA}
\author{B.S.~Acharya} \affiliation{Tata Institute of Fundamental Research, Mumbai-400 005, India}
\author{M.~Adams} \affiliation{University of Illinois at Chicago, Chicago, Illinois 60607, USA}
\author{T.~Adams} \affiliation{Florida State University, Tallahassee, Florida 32306, USA}
\author{J.P.~Agnew} \affiliation{The University of Manchester, Manchester M13 9PL, United Kingdom}
\author{G.D.~Alexeev} \affiliation{Joint Institute for Nuclear Research, Dubna 141980, Russia}
\author{G.~Alkhazov} \affiliation{Petersburg Nuclear Physics Institute, St. Petersburg 188300, Russia}
\author{A.~Alton$^{a}$} \affiliation{University of Michigan, Ann Arbor, Michigan 48109, USA}
\author{A.~Askew} \affiliation{Florida State University, Tallahassee, Florida 32306, USA}
% ...
%
%
\collaboration{The D0 Collaboration\footnote{with visitors from
$^{a}$Augustana College, Sioux Falls, SD 57197, USA,
$^{b}$The University of Liverpool, Liverpool L69 3BX, UK,
$^{c}$Deutshes Elektronen-Synchrotron (DESY), Notkestrasse 85, Germany,
$^{d}$CONACyT, M-03940 Mexico City, Mexico,
$^{e}$SLAC, Menlo Park, CA 94025, USA,
$^{f}$University College London, London WC1E 6BT, UK,
$^{g}$Centro de Investigacion en Computacion - IPN, CP 07738 Mexico City, Mexico,
$^{h}$Universidade Estadual Paulista, S\~ao Paulo, SP 01140, Brazil,
$^{i}$Karlsruher Institut f\"ur Technologie (KIT) - Steinbuch Centre for Computing (SCC),
D-76128 Karlsruhe, Germany,
$^{j}$Office of Science, U.S. Department of Energy, Washington, D.C. 20585, USA,
$^{l}$Kiev Institute for Nuclear Research (KINR), Kyiv 03680, Ukraine,
$^{m}$University of Maryland, College Park, MD 20742, USA,
$^{n}$European Orgnaization for Nuclear Research (CERN), CH-1211 Geneva, Switzerland,
$^{o}$Purdue University, West Lafayette, IN 47907, USA,
$^{p}$Institute of Physics, Belgrade, Belgrade, Serbia,
and
$^{q}$P.N. Lebedev Physical Institute of the Russian Academy of Sciences, 119991, Moscow, Russia.
$^{\ddag}$Deceased.
}} \noaffiliation
\vskip 0.25cm

%% file: acknowledgement_APS_full_names.tex
%

This document was prepared by the D0 collaboration using the resources of the Fermi National Accelerator Laboratory (Fermilab),
a U.S. Department of Energy, Office of Science, HEP User Facility. Fermilab is managed by Fermi Research Alliance, LLC (FRA),
acting under Contract No. DE-AC02-07CH11359.

We thank the staffs at Fermilab and collaborating institutions,
and acknowledge support from the
Department of Energy and National Science Foundation (United States of America);
Alternative Energies and Atomic Energy Commission and
National Center for Scientific Research/National Institute of Nuclear and Particle Physics  (France);
Ministry of Education and Science of the Russian Federation, 
National Research Center ``Kurchatov Institute" of the Russian Federation, and 
Russian Foundation for Basic Research  (Russia);
National Council for the Development of Science and Technology and
Carlos Chagas Filho Foundation for the Support of Research in the State of Rio de Janeiro (Brazil);
Department of Atomic Energy and Department of Science and Technology (India);
Administrative Department of Science, Technology and Innovation (Colombia);
National Council of Science and Technology (Mexico);
National Research Foundation of Korea (Korea);
Foundation for Fundamental Research on Matter (The Netherlands);
Science and Technology Facilities Council and The Royal Society (United Kingdom);
Ministry of Education, Youth and Sports (Czech Republic);
Bundesministerium f\"{u}r Bildung und Forschung (Federal Ministry of Education and Research) and 
Deutsche Forschungsgemeinschaft (German Research Foundation) (Germany);
Science Foundation Ireland (Ireland);
Swedish Research Council (Sweden);
China Academy of Sciences and National Natural Science Foundation of China (China);
and
Ministry of Education and Science of Ukraine (Ukraine).

%% file: d0_zc_v2.bbl
\begin{thebibliography}{99}
\bibitem{eta}
$\eta=-\ln[\tan(\theta/2)]\ $ is the pseudorapidity and $\theta$ is
the polar angle between the track momentum vector and the proton beam direction. $\phi$ is the azimuthal angle 
of the track.
The pseudorapidity defined using the polar angle measured with a coordinate origin at
 the center of the detector is called  $\eta_{\rm det} $.
\bibitem{ip}
 The impact parameter {\sl IP} is defined as the distance of closest
 approach of the track to the $p\bar{p}$
 collision point projected onto the plane transverse to the $p\bar{p}$ beams.
\end{thebibliography}
